\documentclass[pra,twocolumn,floatfix,superscriptaddress,eqsecnum]{revtex4-1}

\usepackage{graphicx} % enhanced support for graphics
\usepackage{color} % use colored text in latex
\usepackage{amsmath} % Defines extra environments for multiline displayed equations, as well as a number of other enhancements for math
\usepackage[urlcolor=blue, hyperindex, colorlinks, bookmarks=true,linkcolor=black,citecolor=black]{hyperref} % provides LaTeX the ability to create hyperlinks within the document
\usepackage{amssymb}% {bm}% bold math
\usepackage{ulem}

\newcommand{\dg}{^\dagger}

\newcommand{\bra}[1]{\langle{#1}|}
\newcommand{\ket}[1]{|{#1}\rangle}

\newcommand{\be}{\begin{equation}}
\newcommand{\ee}{\end{equation}}
\newcommand{\bea}{\begin{eqnarray}}
\newcommand{\eea}{\end{eqnarray}}

\begin{document}
\normalem

\title{Analytic control methods for high fidelity unitary operations in a weakly nonlinear oscillator}
\date{\today}
\author{J. M. Gambetta}
\affiliation{Institute for Quantum Computing and Department of Applied Mathematics, University of Waterloo, Waterloo, Ontario, Canada N2L 3G1}
\author{F. Motzoi}
\affiliation{Institute for Quantum Computing and Department of Physics and Astronomy, University of Waterloo, Waterloo, Ontario, Canada N2L 3G1}
\author{S. T. Merkel}
\affiliation{Institute for Quantum Computing and Department of Physics and Astronomy, University of Waterloo, Waterloo, Ontario, Canada N2L 3G1}
\author{F. K. Wilhelm}
\affiliation{Institute for Quantum Computing and Department of Physics and Astronomy, University of Waterloo, Waterloo, Ontario, Canada N2L 3G1}

\begin{abstract}
In qubits made from a weakly anharmonic oscillator the leading source of error at short gate times is leakage of population out of the two dimensional Hilbert space that forms the qubit. In this paper we develop a general scheme based on an adiabatic expansion to find pulse shapes that correct this type of error. We find a family of solutions that allows tailoring to what is practical to implement for a specific application.  Our result contains and improves the previously developed DRAG technique [F. Motzoi, \emph{et. al.}, Phys. Rev. Lett. \textbf{103}, 110501 (2009)] and allows a generalization to other non-linear oscillators with more than one leakage transition.
\end{abstract}
\maketitle

\section{Introduction}
The physical realization of quantum information processing in
superconducting circuits \cite{Makhlin2001,You2003,Insight,Schoelkopf2008} has enjoyed
remarkable progress over the last decade. While initially decoherence limited single qubits to only a few coherent oscillations \cite{Nakamura1999}, high precision, general quantum conrol is now possible over single- and few-qubit systems.
This is evident by the demonstration of high-fidelity nonclassical states of
two-qubit \cite{Steffen2006,Ansmann2009,DiCarlo2009,Chow2010} and three-qubit \cite{DiCarlo2010,Neeley2010} systems, harmonic oscillators \cite{Hofheinz2009}, and the demonstration of small quantum algorithms \cite{DiCarlo2009}.

This success is partially due to our current understanding of sources of noise and the development of techniques and systems that are  resilient to these noise sources. Examples include the optimum working point \cite{Vion2002} and the introduction of low-dispersion qubits like the transmon \cite{Koch2007,Schreier2008} and the capacitively shunted flux
qubit \cite{Steffen2010}. On the other hand, a promising route to success are qubits that contains only a minimal number of elements, such as the phase qubit  \cite{Martinis2002,Steffen2006a,Martinis09c}. What these systems have in common is a weakly anharmonic energy level structure, i.e., the states that are outside of the qubit subspace spanned by $|0\rangle$ and $|1\rangle$ are only separated from each other and the qubit subspace by energies only slightly different than the qubit frequency. 

Having a weakly anharmonic qubit poses a challenge in the implementation of quantum gates. It is known \cite{Husimi53,Feynman65} that the time evolution operator of a linearly driven harmonic oscillator is a combination of a coherent displacement operator tracking the classical trajectory
of the driven oscillator and a global phase factor. This evolution encompasses all energy levels and cannot be reduced to a single-qubit rotation. Spectroscopically, this can be understood as follows: a single qubit rotation is typically implemented by a pulse of radiation resonant with the qubit energy splitting.
In a harmonic oscillator, all energy splittings are the same, so driving one transition drives all others at the same time. A system starting initially in an energy eigenstate will quickly be driven into a superposition over many energy eigenstates. By this token, it is crucial that a qubit is nonlinear \cite{Vion2002,Makhlin2001,Schoelkopf2008,Insight,Khani2009}, that is that the transition frequency of the qubit levels is different by an amount $\Delta$ from the transition frequencies to the non-qubit levels. Spectroscopically, we would expect that whenever the bandwidth of the pulse comes close to $\Delta$, i.e., when its duration becomes short on the scale of $1/\Delta$, we expect
significant leakage to higher states. Thus, it is a challenge to implement fast single-qubit gates in weakly anharmonic systems. The implementation of faster gates is important as it allows more gates to be executed in a given coherence time, an important step toward high-fidelity quantum logic.

While superconducting qubits are the most well known example of qubits made from weakly anharmonic oscillators, there are many other examples. In fact, it has been shown that no physical particle can be a true qubit\cite{Wu02}. Examples of leakage states include higher vibrational states in optical lattices \cite{Maneshi08}, polarized spin states in the singlet-triplet qubits \cite{Petta05}, and auxiliary states in ion traps \cite{Haeffner08} and Rydberg atoms \cite{Urban09,Gaetan09,Haroche2006}.

In this paper, we outline a suite of strategies to implement single-qubit quantum gates in qubits singled out from the spectrum of an anharmonic oscillator.
We develop an adiabatic expansion technique that leads to order-by-order constraint equations on a toggling-frame transformation and the control fields. The space of solutions contains the DRAG strategy proposed in Ref. \cite{Motzoi2009} and re-analyzed in the presence of an oscillator bath in Ref. \cite{Poudel10} as a special case, as well as simpler versions of DRAG that do not require time-dependent energy bias or phase ramping, similar to those
implemented in Refs. \cite{Chow2010a} and \cite{Lucero2010}. Additionally, we derive optimal solutions to a given order by minimizing the errors manifested by the fields in the next higher order. 

The outline of the paper is as follows: in Sec. \ref{sec:system} we review qubits made from anharmonic oscillator and introduce their rotating-wave description. Sec. \ref{sec:error} describes Gaussian pulses and shows how they lead to both phase and population errors. In Sec. \ref{sec:expansion} we extend the DRAG scheme of Ref. \cite{Motzoi2009} to find a wealth of different pulse shapes that give improved performance over Gaussian shaping. In Sec. \ref{sec:examples} we apply the generalized scheme to cases where there are more then one leakage level. Finally we conclude in Sec. \ref{sec:conclude}.

\section{The system \label{sec:system}}

\subsection{Lab frame Hamiltonian}

\begin{figure}[htbp]
\centering
\includegraphics[width=0.20\textwidth]{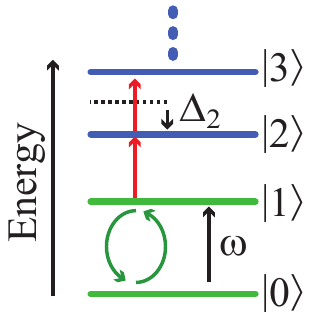}
\caption{(color online) Energy level diagram of the system we are considering. The qubit is formed by the $\ket{0}$ and $\ket{1}$ (green) levels and we aim to have complete control in this subspace when leakage to the $\ket{2}$ and then $\ket{3}$ etc.~is possible (red arrows). The dotted black lines indicate the positions the energy levels would be at if the system was a harmonic oscillator of frequency $\omega$.}\label{fig:Ham}
\end{figure}

We consider a qubit formed by the two lowest levels (which we generalize in sec.~\ref{sec:examples}) of an anharmonic oscillator. These levels are separated in energy by $\hbar\omega$, where $\omega$ is the transition frequency. The $j^\mathrm{th}$ higher levels are different to $\hbar j\omega$ by $\hbar \Delta_j$, where $\Delta_j$ is known as the anharmonicity. That is, the Hamiltonian for the nonlinear oscillator of dimension $d$ is ($\hbar=1$)
\begin{equation}
    H_\mathrm{fr}=\sum_{j=1}^{d-1}(j\omega+\Delta_j)\Pi_j,
\label{eq:hamiltonian}\end{equation} where $\Pi_j=\ket{j}\bra{j}$ is the projection operator onto the $j^\mathrm{th}$ energy level. Without loss of generality we set $\Delta_1=0$. This is illustrated in Fig. \ref{fig:Ham} with the qubit levels being $\ket{0}$ and $\ket{1}$ (green). 
 For many nonlinear oscillators the anharmonicity takes the form $\Delta_j=\Delta_2(j-1)j/2$, which we will call the standard nonlinear oscillator (SNO), essentially a Duffing oscillator within the rotating wave approximation \cite{Khani2009}. For the lowest few levels, superconducting qubits of the transmon \cite{Schreier2008}, phase qubit \cite{Martinis09c}, and capacitively shunted flux qubit  \cite{Steffen2010} types are well approximated by a SNO. Furthermore,  motional states in optical lattices \cite{Maneshi08}, collective modes of ion traps \cite{Haeffner08} and nanomechanical oscillators \cite{Kozinsky07} are also described as SNOs.

%\begin{table*}
%\centering
%	\begin{tabular}{l|cccccc}
%	\hline
%\hline
%   &\scriptsize{Transmon\cite{Schreier2008}}  & \scriptsize{Phase qubit\cite{Martinis2005}}& \scriptsize{CSFQ\cite{Steffen2010}} & \scriptsize{Other}& \scriptsize{Other} & \scriptsize{Other}\\ \hline
%\scriptsize{$\omega/2\pi$} & \scriptsize{$6\rightarrow10$ GHz} & \scriptsize{$1\rightarrow10$ GHz} & \scriptsize{$1\rightarrow10$ GHz} & \scriptsize{X} \\
%\scriptsize{$\Delta/2\pi$} & \scriptsize{$-200 \rightarrow -400$ MHz} & \scriptsize{$-200 \rightarrow -400$ MHz} & \scriptsize{$200 \rightarrow 400$ MHz} & \scriptsize{X} \\
%\scriptsize{$\alpha$} & \scriptsize{$1/10$} & \scriptsize{$1/10$} & \scriptsize{$1/10$} & \scriptsize{$1/10$} \\
%\hline\hline
%\end{tabular}
%\caption{\label{tab:val}Typical values for some typical nonlinear oscillators that are well described by our theory.\jcomment{table needs doing including reviewing the superconducting qubit parameters.}}
%\end{table*}

We will assume that control in this system is due to some dipole-like interaction that only allows single photon transitions. As for harmonic oscillators, this is a good approximation because parity forbids all other transitions. The control Hamiltonian is
\begin{equation}\label{eq:controlham}
    H_\mathrm{ct}(t)=\mathcal{E}(t)\sum_{j=1}^{d-1}\lambda_{j-1}\sigma^x_{j-1,j},
\end{equation} where $\mathcal{E}(t)$ is the drive amplitude,  $\sigma^x_{j,k}=\ket{j}\bra{k}+\ket{k}\bra{j}$ is one of the effective Pauli spin operators for levels $j$ and $k$, and $\lambda_j$ is a dimensionless parameter that weighs the relative strength of driving the $\ket{j}\rightarrow \ket{j+1}$ transition versus the $\ket{0}\rightarrow \ket{1}$ transition.  In our model we take $\lambda_0=1$ and leave the $\lambda_j$'s as input parameters. For a harmonic oscillator controlled via a dipole interaction with an external field $\lambda_j=\sqrt{j}$;
however, in appendix \ref{sec:lamb} we show that in cavity or circuit QED architectures $\lambda_j$ can differ substantially from this value.

For the functional form of the drive $\mathcal{E}(t)$ we will assume that  $\left|\mathcal{E}(t)\right| \ll \omega$ (weak driving regime) and introduce envelope shaping of the driving field at carrier frequency $\omega_d$.  This leads to
\begin{equation}\label{eq:controls}
\mathcal{E}(t)=\Omega_x(t)\cos(\omega_d t + \phi_0)+\Omega_y(t)\sin(\omega_d t + \phi_0).
\end{equation}
As per convention, the two quadratures amplitudes $\Omega_x(t)$ and $\Omega_y(t)$ can be amplitude modulated using a waveform generator and then mixed back together with the carrier to give this form of control field. Here,
$\phi_0$ is the relative phase between the envelope and the carrier at the start of the operation. This phase is irrelevant if the rotating wave approximation can be made, as will be shown in the next section.

\subsection{Rotating frame Hamiltonian}\label{sec:Rot}

For quantum information processing it is highly suitable to define operations with respect to the frame rotating at the  driving frequency  $\omega_d$. In this frame we have three independent controls: $\delta(t) = \omega(t)-\omega_d$ (the qubit detuning), $\Omega_x(t)$ and $\Omega_y(t)$, which, projected to the qubit subspace, control application of the three Pauli spin operators $\sigma_{0,1}^z$, $\sigma_{0,1}^x$, and $\sigma_{0,1}^y$, respectively. For example, the identity operation is achieved by setting all the controls to zero. Note that here we have assumed that $\delta$ is controlled by shifting the qubit frequency. This is not necessary and, as shown in appendix \ref{sec:det}, this can be achieved by a time dependent phase in Eq. \eqref{eq:controls}. To move to the rotating frame we define the unitary 
\begin{equation}\label{eq:rot}
 R(t)=\sum_{j=1}^{d-1}\exp\left[-i j\omega_d t \right]\Pi_j,
\end{equation} which determines the transformed Hamiltonian
\begin{equation}\label{eq:rotransf}
 H^R(t)=R^\dagger(t)  H(t) R(t)+i\dot R^\dagger(t) R(t),
\end{equation} where $H(t)=H_\mathrm{fr}+H_\mathrm{ct}(t)$. Explicitly we have
\begin{equation}
 \begin{split}
H^R(t)=&\sum_{j=1}^{d-1}(j\delta+\Delta_j)\Pi_j + \left[\frac{\Omega(t)}{2}e^{-i\omega_d t-i\phi_0}+\mathrm{h.c.}\right]\\&\times\sum_{j=1}^{d-1}\lambda_{j-1}\left[\ket{j}\bra{j-1}e^{i\omega_d t}+\mathrm{h.c.}\right],\end{split}
\end{equation} where h.c. stands for hermitian conjugate and $\Omega(t) = \Omega_x(t)+i\Omega_y(t)$.

Assuming that $\omega_d$ is larger then any other rate or frequency in this frame we can perform the rotating wave approximation (i.e.~time average the fast rotating terms to zero). For the SNO case this amounts to restricting the dimension $d$ to be less then $\sqrt{2\omega/\Delta}$, specifically $d= 7$. After this approximation we can write the Hamiltonian as
\begin{eqnarray}\label{eq:rotHamstep}
H^R(t)&=&\sum_{j=1}^{d-1}(j\delta+\Delta_j)\Pi_j\\
&\ & + \ \sum_{j=1}^{d-1}\lambda_{j-1}\left[\frac{\Omega(t)}{2}\ket{j}\bra{j-1}+\mathrm{h.c}\right].\nonumber
\end{eqnarray} 
Here we have included the $\phi_0$ into the energy states ($\ket{j}\rightarrow e^{ij\phi_0}\ket{j}$), and we see that within the rotating wave approximation the relative phase between the envelope and the carrier at the start of the operation is irrelevant.
Finally, Eq. \eqref{eq:rotHamstep} can be rewritten as
 \begin{equation}\label{eq:rotHam}
\begin{split} H^R(t)=&\sum_{j=1}^{d-1}(j\delta(t)+\Delta_j)\Pi_j \\&+\sum_{j=1}^{d-1}\lambda_{j-1}\left[\frac{\Omega_x(t)}{2}\sigma_{j-1,j}^x +\frac{\Omega_y(t)}{2}\sigma_{j-1,j}^y\right],
\end{split}
\end{equation} where $\sigma_{j,k}^y=-i\ket{j}\bra{k}+i\ket{k}\bra{j}$ for $k>j$.  We see that if we can restrict the system to the lowest two levels then all rotations in the single qubit space can be achieved by independent controls; however, in general, this is not true. In Sec.~\ref{sec:error} we show that the higher level transitions lead to a combination of a phase and leakage error \cite{Steffen2003,Motzoi2009}. This has been experimentally measured in Refs. \cite{Chow2009} and \cite{Lucero2008}. 

\section{Gaussian shaping and errors}\label{sec:error}

Our goal is to implement gates contained within the qubit subspace. That is, we want to shape $\Omega_x(t)$, $\Omega_y(t)$ and $\delta(t)$ in Eq. \eqref{eq:rotHam}  so that
\begin{equation}\label{eq:ideal}
	U_\mathrm{ideal} = \mathcal{T}\exp\left[-i\int_{0}^{t_g}H^R(t)dt\right] = e^{i\phi}U_\mathrm{qb}\oplus U_{\mathrm{rest}},
\end{equation}  where $t_g$ is the gate time, $\mathcal{T}$ is the time ordering operator, $U_\mathrm{qb}$ is a unitary that acts only in the qubit subspace, $U_{\mathrm{rest}}$ acts only outside of the qubit space, and $\phi$ describes a relative phase.  Therefore, $U_{\mathrm{rest}}$ as well as the phase $\phi$ are  completely irrelevant for operations in the Hilbert space formed by the qubit. 

\begin{figure}[htbp]
\centering
\includegraphics[width=0.45\textwidth]{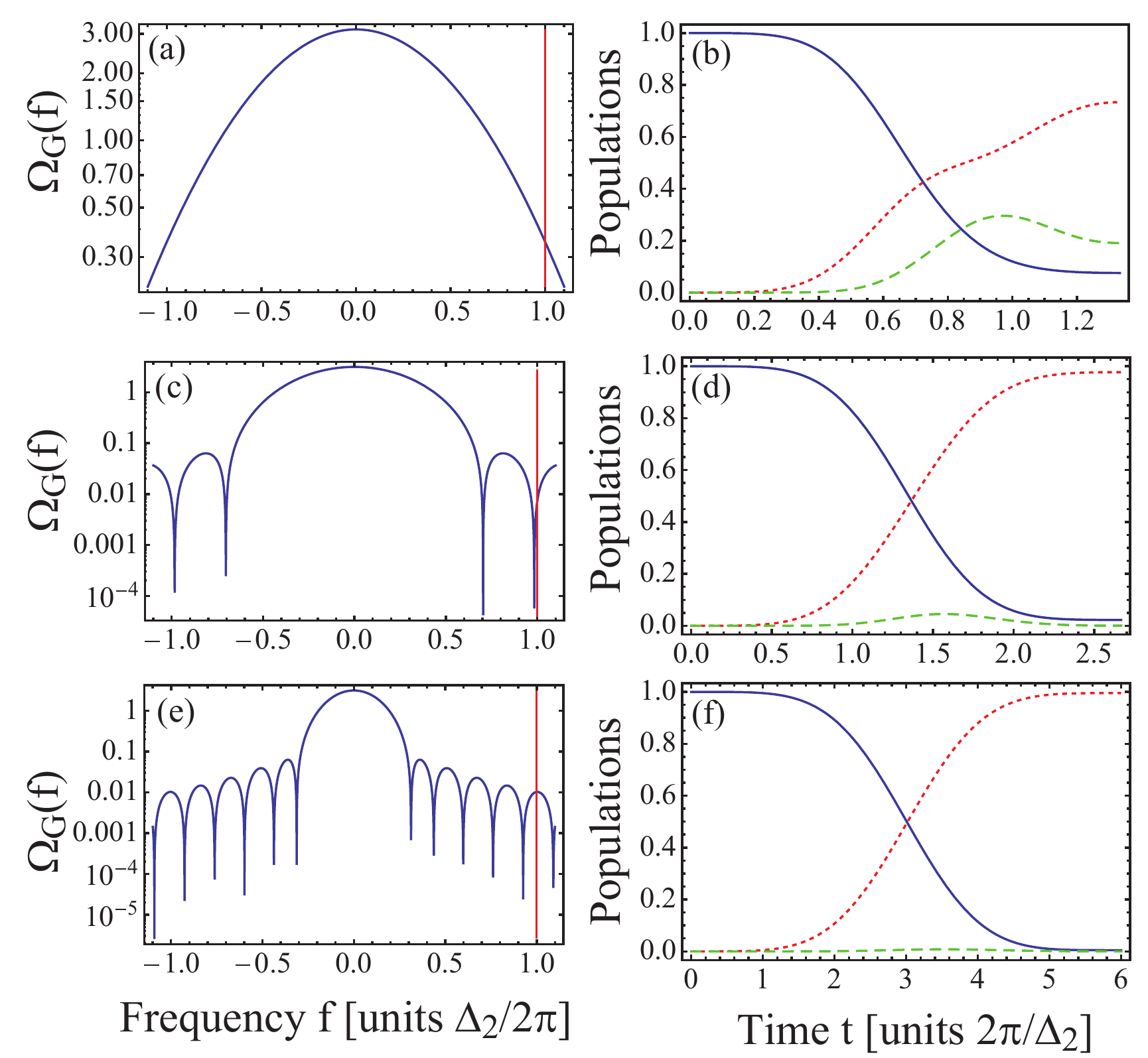}
\caption{(color online) Fourier transforms of the control fields (a, c, e) and populations (b, d, f) of the ground (blue solid), first (red dotted), and second excited (green dash) in a simulation of a NOT gate with a Gaussian amplitude pulse for a $d=5$ SNO. In (a, b) $\sigma=2\pi/3\Delta_2$, (c, d) $\sigma=4\pi/3\Delta_2$, and (e, f) $\sigma=3\pi/\Delta_2$ and the gate time is taken to be $4\sigma$.}\label{fig:Guas}
\end{figure}

 To demonstrate the typical set of errors we choose $U_\mathrm{qb} =\sigma_{0,1}^x$, the NOT gate. For a leakage-free qubit this would be implemented by simply setting $\delta(t)=0$, ${\Omega}_x(t)=\Omega_G(t)$, and ${\Omega}_y(t)=0$, with the only requirement that that $\int_{0}^{t_g}{\Omega}_G (t)dt=\pi$. To reduce the leakage to the third level, the standard result prior to Ref. \cite{Motzoi2009} was to use Gaussian modulation of the envelope  \cite{Bauer1984,Steffen2003}. In this case $\Omega_G(t)$ takes the form
\begin{equation}\label{eq:gaussian}
\Omega_G(t)=
			A\frac{\exp\left[-\frac{(t - t_g/2)^2}{2\sigma^2}\right]-\exp\left[-\frac{t_g^2}{8\sigma^2}\right]}{\sqrt{2\pi \sigma^2}\mathrm{erf}[t_g/\sqrt{8}\sigma]-t_g\exp[-t_g^2/8\sigma^2]},
\end{equation} 
where $t\in [0,t_g]$,  $\sigma$ is the standard deviation and $A$ is chosen such that the correct amount of rotation is implemented (e.g. $A=\pi$ for a NOT). This functional form is chosen to enforce that the pulse start and end at zero. In the limit that $t_g\rightarrow\infty$ we recover the standard Gaussian function. The motivation for Gaussian shaping is that the small and strictly limited frequency bandwidth  $(1/\sigma)$ ensures little excitation at the leakage transition frequency. For short pulses however, there is still significant spectral weight at $\Delta_2$. This is shown in Fig. \ref{fig:Guas} (a, c, e) where the Fourier transforms of $\Omega_G(t)$ are plotted for $\sigma =\{1/3,2/3,3/2\}2\pi/\Delta_2$ and $t_g=4\sigma$ respectively.

To quantify this error, we use the gate fidelity averaged over all input states existing in the qubit Hilbert space,
\begin{equation}
	\begin{split}	
		F_g	=&\int d\psi ~\mathrm{Tr}\left[U_\mathrm{ideal} \ket{\psi}\bra{\psi} U_\mathrm{ideal}\dg \mathcal{E}(\ket{\psi}\bra{\psi})\right],
	\end{split}
\end{equation} where $\mathcal{E}(\rho)$ is the actual process in the full Hilbert space. Using an argument similar to Ref. \cite{Bowdrey2002} gives
\begin{equation}
	\begin{split}\label{eq:fid}	
		F_g	=&\frac{1}{6}\sum_{j=\pm x, \pm y, \pm z}\mathrm{Tr}\left[U_\mathrm{ideal} \rho_j U_\mathrm{ideal}\dg \mathcal{E}(\rho_j)\right],
	\end{split}
\end{equation}
where $\rho_j$ are the six axial states on the Bloch sphere, and $U_\mathrm{ideal}$ is defined in Eq. \eqref{eq:ideal}. To demonstrate the errors arising from Gaussian shaping we consider a $d=5$ SNO and numerically calculate the gate error ($1-F_g$) for $\sigma =\{1/3,2/3,3/2\}2\pi/\Delta_2$ and $t_g=4\sigma$. We find gate errors of $0.198$, $0.0160$, and $0.0030$ respectively.

To understand these error values we plot the populations of the first three levels in Fig. \ref{fig:Guas} (b, d, f). The ground state populations are given by the blue solid line; the red dotted line shows the first excited state; the green dashed line is the second excited state. We observe that for the shortest gate, $\sigma=2\pi/3\Delta_2$,  the error after the pulse is mostly residual population of the third and higher level. This is what we refer to as the leakage error. For longer gates, e.g. $\sigma=4\pi/3\Delta_2$, the residual population does not account for the calculated error. This error is mostly a phase error resulting from the finite population of the third level {\em during} the pulse. Even though the final state is restricted to the computational levels, the admixture of the third level leads to a phase shift on the second level, resulting in a net phase error at the end of the pulse. At the longest time when the population of the third level is nearly negligible there is still a large gate error. From these results we conclude that Gaussian shaping is of limited performance even if the pulse bandwidth is somewhat smaller than $\Delta_2$. Thus a more advanced pulse is needed. In Ref. \cite{Motzoi2009}, we provided a simple scheme and in the next section we will review this and generalize the result.

\section{Simple adiabatic control pulses}\label{sec:expansion}

\subsection{General procedure for DRAG solutions}

To go beyond the Gaussian control methods presented above, we introduce the Derivative Removal by Adiabatic Gate (DRAG) technique, generalizing the result of Ref. \cite{Motzoi2009}.
In this technique we want to find a time-dependent unitary transformation $A(t)$ such that the effective Hamiltonian
\begin{equation}\label{eq:Heff}
H_\mathrm{eff}(t) = A^\dagger (t) H(t) A(t) + i \dot{A}^{\dagger}(t) A(t).
\end{equation} has the form $H_\mathrm{qb}(t)\oplus H_\mathrm{rest}(t)$ where $H_\mathrm{qb}(t)$ is a Hamiltonian in the effective qubit subspace and $ H_\mathrm{rest}(t)$ generates evolution in the rest of the system. The direct sum form implies that if the state of the system starts in the qubit subspace, then it remains there during and after the pulse. We impose the  additional requirement that $A(0) = A(t_g) = \openone$, i.e. the frame transformation vanishes at the boundaries. Both conditions together imply that we have decoupled the computational subspace from the leakage subspace since the qubit subspace is mapped back onto itself by the end of the pulse.  In Ref. \cite{Motzoi2009} we restricted the problem to a $d=3$ system and found a solution to the problem of generating a NOT gate. It is straightforward to generalize this work to design pulses that account for higher order corrections, larger dimensional embeddings, general target gates, and an expanded library of control fields by considering more complicated frame transformations to that considered in Ref. \cite{Motzoi2009}. We now classify valid choices for the transformation $A(t)$ that satisfies the above constraints. 

To simplify the following arguments we consider a basis for the Lie algebra $\mathfrak{u}(d)$, as opposed to $\mathfrak{su}(d)$, with elements $\{\sigma_{j,k}^x, \sigma_{j,k}^y, \Pi_l \}$.  Here $1\leq l\leq d-1$ and $0\leq j<k\leq d$.
With respect to this basis we wish to implement a Hamiltonian in the qubit subspace of the form
\begin{equation}
H_\mathrm{qb}(t)= {\frac{1}{2}}\left[h_x(t)\sigma^x_{0,1}+
h_y(t)\sigma^y_{0,1}\right] + \frac{1}{2}h_z(t) \left(\Pi_0 - \Pi_1\right) .
\label{eq:Hwantq}\end{equation} 
Physically this amounts to setting
conditions  on the control fields and the frame transformation such that
\begin{eqnarray}\label{eq:conds}
&&\hspace{-5mm}\mathrm{Tr}[ H_\mathrm{eff}(t) \sigma^x_{0,1}]=h_x(t),  \\
&&\hspace{-5mm}\mathrm{Tr}[ H_\mathrm{eff}(t) \sigma^y_{0,1}]= h_y(t),  \\
&&\hspace{-5mm}\mathrm{Tr}[ H_\mathrm{eff}(t)  \left(\Pi_0 - \Pi_1 \right)]= h_z(t), \label{eq:conds2}\\
&&\hspace{-5mm}\mathrm{Tr}[ H_\mathrm{eff}(t) \sigma^x_{0,k}]=0~  \mathrm{for}~2 \leq k \leq d-1,\label{eq:conds2a}\\
&&\hspace{-5mm}\mathrm{Tr}[ H_\mathrm{eff}(t) \sigma^y_{0,k}]=0~  \mathrm{for}~2 \leq k \leq d-1,\\
&&\hspace{-5mm}\mathrm{Tr}[ H_\mathrm{eff}(t) \sigma^x_{1,k}]=0~  \mathrm{for}  ~2 \leq k \leq d-1,\\
&&\hspace{-5mm}\mathrm{Tr}[ H_\mathrm{eff}(t) \sigma^y_{1,k}]=0~  \mathrm{for}~  2 \leq k \leq d-1.\label{eq:conds3}
\end{eqnarray}
The three first equations define the controls in the qubit subspace whereas the last four impose the condition of no leakage. There are no further conditionsso that the dynamics strictly inside the leakage subspace remains arbitrary.

To find a scheme that satisfies Eqs. \eqref{eq:conds} -- \eqref{eq:conds3} we write $A(t)=\exp[-i S(t)]$ where $S(t)$ is an arbitrary Hermitian operator that we decompose as
\begin{equation}
S(t) = \sum_{j=1} s_{z,j}(t) \Pi_j + \sum_{j<k} s_{x,j,k}(t) \sigma^x_{j,k} +  \sum_{j<k}s_{y,j,k}(t) \sigma^y_{j,k},
\end{equation} and by assuming a power series in a small parameter $\epsilon$, we can write each element as
\be
s_{\alpha,j,k}(t) = \sum_{n=1}^{\infty} s^{(n)}_{\alpha,j,k}(t) \epsilon^l,
\ee where $\alpha =x,y,$ or $z$. This ensures that the transformation is perturbative with respect to the parameter $\epsilon$. We take $\epsilon=1/t_g\Delta$ as a small parameter, implying that, for the fast gates we are interested in, $\Omega_G/\Delta$ will be of order $\epsilon$. Furthermore, since we are interested in unitary operations, we will define a dimensionless time and Hamiltonian whereby $\int_0^{t_g} H(t)dt =\int_0^1 \bar H(t)dt$ (so $\bar H(t)=t_g H$).  Doing this allows us to write the Hamiltonian for the system in Eq. \eqref{eq:rotHam} as
\begin{equation}\label{eq:dimHam}
 \bar H(t) =\frac{1}{\epsilon} H_0 + \bar\delta(t) H_{z} + \frac{\bar\Omega_{x}(t)}{2} H_{x}+ \frac{\bar\Omega_{y}(t)}{2}  H_{y},
\end{equation} where
\begin{eqnarray}
H_0 &=& \sum_{j=2}^{d-1} \frac{\Delta_j}{\Delta_2}\Pi_j,\\
H_z &=& \sum_{j=1}^{d-1}j \Pi_j,\\
H_x &=& \sum_{j=1}^{d-1} \lambda_{j-1} \sigma_{j-1,j}^x, \\
H_y &=& \sum_{j=1}^{d-1} \lambda_{j-1} \sigma_{j-1,j}^y.
\end{eqnarray} Here $\bar\delta(t)$, $\bar\Omega_{x}(t)$, and $\bar\Omega_{y}(t)$ are dimensionless versions of the previously defined matrix elements scaled by the rule $\bar\delta(t)=t_g \delta(t)$. Note we have dropped the $R$ superscript and assume from now on we are in the rotating frame.

For the control fields we also write a series expansions
\begin{eqnarray}
\bar\Omega_{x}(t)&=& \sum_{n=0}^{\infty} \epsilon^n \bar\Omega^{(n)}_{x}(t), \\
\bar\Omega_{y}(t)&=& \sum_{n=0}^{\infty} \epsilon^n \bar\Omega^{(n)}_{y}(t), \\
 \bar\delta(t)&=& \sum_{n=0}^{\infty} \epsilon^n \bar\delta^{(n)}(t),
\end{eqnarray} 
 and by the results presented in appendix \ref{A:general}, the constraints Eqs. \eqref{eq:conds} -- \eqref{eq:conds2} can be rewritten as
\begin{eqnarray} \label{eq:controlcon}
   \bar\Omega_{x}^{(n)}(t) & =& h_x^{(n)}(t)-\mathrm{Tr} [ {H}_\mathrm{extra}^{(n)}(t) \sigma^x_{0,1} ], \\ç¬
 \bar\Omega_{y}^{(n)}(t) &=& h_y^{(n)}(t)- \mathrm{Tr} [ {H}_\mathrm{extra}^{(n)}(t) \sigma^y_{0,1} ], \\
 \bar\delta^{(n)}(t) & =& \mathrm{Tr} [ {H}_\mathrm{extra}^{(n)}(t)\left(\Pi_0 - \Pi_1 \right)]-h_z^{(n)}(t), \label{eq:controlcon2}
\end{eqnarray} where ${H}_\mathrm{extra}^{(n)}(t)$ contains terms generated by the lower orders of the transformation and the controls.
It is a rather complicated expression which can be derived following the procedure in appendix \ref{A:general} and the first few orders are listed in Eqs. \eqref{eq:hextra} -- \eqref{eq:hextra3}.  For the leakage constraints Eqs. \eqref{eq:conds2a} -- \eqref{eq:conds3}, we also derive in appendix \ref{A:general} the following constraints on the frame transformation: 
\begin{eqnarray}
\label{eq:framecon}
 s^{(n+1)}_{y,0,k}(t)&=&-\frac{\Delta_2}{2 \Delta_k}\mathrm{Tr}[{H}_\mathrm{extra}^{(n)}(t) \sigma^x_{0,k}],\\
s^{(n+1)}_{x,0,k}(t)&=&\frac{\Delta_2}{2 \Delta_k}\mathrm{Tr}[{H}_\mathrm{extra}^{(n)} (t)\sigma^y_{0,k}]\label{eq:framecon2b},\\
 s^{(n+1)}_{y,1,k}(t)&=&\frac{\Delta_2}{2 \Delta_k}\mathrm{Tr}[{H}_\mathrm{extra}^{(n)} (t)\sigma^x_{1,k}] -  \frac{\bar\Omega_x^{(n)}(t)\lambda_1\delta_{k,2}}{2}, \label{eq:framecon2a}\nonumber\\\\
s^{(n+1)}_{x,1,k}(t)&=&\frac{\Delta_2}{2 \Delta_k}\mathrm{Tr}[{H}_\mathrm{extra}^{(n)}(t)\sigma^y_{1,k}] + \frac{\bar\Omega_y^{(n)}(t)\lambda_1\delta_{k,2}}{2}.\nonumber\\
 \label{eq:framecon2}
\end{eqnarray}
The coefficients $s^{(n+1)}_{x,0,1}(t)$, $s^{(n+1)}_{y,0,1}(t)$, and $s^{(n+1)}_{z,i}(t)$ are free parameters in our theory.  Choosing a different functional form for them result in different DRAG solutions. In the next section we will give some practical examples for this choice.

\subsection{Zero, first, and second order DRAG solutions}

\subsubsection{Zero order solution}

\begin{figure}[htbp]
\centering
\includegraphics[width=0.40\textwidth]{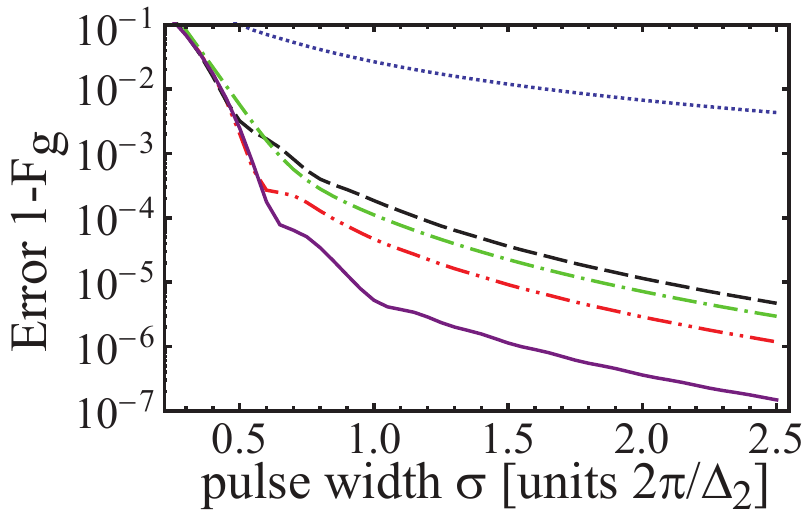}
\caption{(color online) Gate error for the implementation of a NOT gate in a $d=5$ SNO as a function of $\sigma$, with $t_g=4\sigma$, and a Gaussian shaped pulse. The blue dotted line is the zeroth order solution. The black dash line is the first order $Z$-only correction. The red dash-dot-dot line is the first order $Y$-only correction. The green dash-dot line is the first order correction from the controls presented in Ref. \cite{Motzoi2009}. The purple solid line is for the optimal first order correction.}\label{fig:DragFig3}
\end{figure}

For definiteness we choose the target Hamiltonian to be $\Omega_{G}(t) \sigma^x_{01}/2$ corresponding to rotations around the $x$-axis. Rotations around the $y$-axis will follow a similar procedure, and $z$-axis rotations are trivial. For this target Hamiltonian we require $h_x(t)=t_g\Omega_G(t)$, $h_y(t)=0$, and $h_z(t)=0$. For simplicity, we take $h_x^{(0)}(t)=t_g\Omega_G(t)$ and  $h_x^{(n)}(t)=0$ for $n>0$, and we take $h_y^{(n)}(t)=h_z^{(n)}(t)=0$ for $n\geq0$. From Eq. \eqref{eq:hextra}, the zeroth order expression $H_\mathrm{extra}^{(0)}(t)$ is zero. This implies that the control constraints in Eqs. \eqref{eq:controlcon} -- \eqref{eq:controlcon2}, for order zero, are
\begin{equation}
\begin{split}
 \label{eq:zerodim}
\bar\Omega_{x}^{(0)}  (t)=   t_g\Omega_G(t) ,\quad \bar\Omega_{y}^{(0)} (t)= 0, \quad \bar\delta^{(0)}  (t)= 0,
\end{split}
\end{equation} giving control solutions
\begin{equation}
\begin{split}\label{eq:zero}
\Omega_{x} (t)=  \Omega_G(t) ,\quad \Omega_{y} (t)= 0,\quad \delta  (t)= 0.
\end{split}
\end{equation} These are the controls used in Sec. \ref{sec:error} and here we will use them as a benchmark for the higher order solutions. In Fig. \ref{fig:DragFig3}, we plot the error, $1-F_g$  with $F_g$ given by Eq. \eqref{eq:fid} (blue dashed line), between a NOT gate an a unitary from the control field given by Eq. \eqref{eq:gaussian} with $A=\pi$ and $t_g=4\sigma$ for a SNO with $d=5$. In this figure it is clearly seen that the error associated with these controls is quite large; for fast gate times this error is unacceptable for quantum information processing, and long gate times will have additional error arising from decoherence.

\subsubsection{First order solution}

To determine the first order solutions we need $H_\mathrm{extra}^{(1)}(t)$, which requires determining the frame transformation conditions for $S^{(1)}(t)$. From Eqs. \eqref{eq:framecon} -- \eqref{eq:framecon2}, these are
\begin{equation}
\begin{split}
  s^{(1)}_{x,0,k}(t) = &0,\quad s^{(1)}_{y,0,k}(t) =0,\\ s^{(1)}_{x,1,k}(t) = &0,\quad s^{(1)}_{y,1,k} (t)= -\lambda_1 t_g\Omega_{G}(t)\delta_{k,2}/2.
\end{split}
\end{equation}
 Using Eq. \eqref{eq:hextra1} for $H_\mathrm{extra}^{(1)}(t)$, the first order corrections to the control fields are
\begin{eqnarray}\label{eq:controlfeilds}
\bar\Omega_{x}^{(1)} (t) &=&  2\dot{s}^{(1)}_{x,0,1} (t) ,\\
 \bar\Omega_{y}^{(1)}(t) &=&   2 \dot{s}^{(1)}_{y,0,1} (t)-s^{(1)}_{z,1} (t)t_g\Omega_G (t),\\
  \bar\delta^{(1)} (t) &=&\dot{s}^{(1)}_{z,1}(t) + 2 s^{(1)}_{y,0,1}(t)t_g\Omega_G(t) + \frac{\lambda_1^2 t_g^2\Omega_G^2(t)}{4}.\nonumber\\
\end{eqnarray} 
Here we see that there is a continuous family of DRAG pulses; however, in this section we will consider four particular solutions. In all of these solutions we take ${s}^{(1)}_{x,0,1} (t)=0$ as it has no influence on our choice for $\bar\Omega_{y}^{(1)}(t)$ and $\bar\delta^{(1)} (t)$.

The first solution we consider is one where the control field $\Omega_{y}(t)=0$. This is achieved by setting $s^{(1)}_{z,1} (t)= 2 \dot{s}^{(1)}_{y,0,1} (t)/t_g\Omega_G (t)$, resulting in
\begin{equation}
\begin{split}
\bar\delta^{(1)} (t) = &\frac{2 \ddot{s}^{(1)}_{y,0,1} (t)}{t_g\Omega_G (t)} - \frac{2 \dot{s}^{(1)}_{y,0,1} (t)\dot\Omega_G (t)}{t_g\Omega_G^2 (t)}\\&+ 2 s^{(1)}_{y,0,1}(t)t_g\Omega_G(t) + \frac{\lambda_1^2 t_g^2\Omega_G^2(t)}{4}.
  \end{split}
\end{equation} The simplest solution that satisfies $S^{(1)}(0)=S^{(1)}(t_g)=0$ is $s^{(1)}_{y,0,1}(t)=0$. In this case the controls become
\begin{equation}
\begin{split}
\Omega_{x}(t) =&  \Omega_G ,\quad
\Omega_{y}(t) =  0, \quad
\delta(t) = \frac{\lambda_1^2 \Omega_G^2(t)}{4\Delta_2}.
  \end{split}
\end{equation} For the SNO considered in Fig. \ref{fig:DragFig3}, the error for this control set is plotted as the black dashed line. It clearly has a much lower error then the standard Gaussian amplitude modulation control, and we will refer to this as the $Z$-only correction.

The second control solution we consider is when the control field $\delta(t)=0$. This is achieved by setting $s^{(1)}_{y,0,1}(t)=-\dot{s}^{(1)}_{z,1}(t)/2t_g\Omega_G(t) -{\lambda_1^2 t_g\Omega_G(t)}/{8}$, which results in
\begin{equation}
\begin{split}
 \bar\Omega_{y}^{(1)}(t) =& -\frac{\ddot{s}^{(1)}_{z,1}(t)}{t_g\Omega_G(t)}+\frac{\dot{s}^{(1)}_{z,1}(t)\dot\Omega_G(t)}{t_g\Omega^2_G(t)}\\&  -\frac{\lambda_1^2 t_g\dot\Omega_G(t)}{4}-s^{(1)}_{z,1} (t)t_g\Omega_G (t).
  \end{split}
\end{equation} Again, the simplest solution that satisfies $S^{(1)}(0)=S^{(1)}(t_g)=0$ is $s^{(1)}_{z,1}(t)=0$. In this case, the controls become
\begin{equation}
\begin{split}
\Omega_{x}(t) = &  \Omega_G(t) ,\quad
\Omega_{y}(t) =  -\frac{\lambda_1^2 \dot\Omega_G(t)}{4\Delta_2},\quad
\delta(t) =  0,
  \end{split}
\end{equation} and, for the SNO considered in the numerical demonstration, the error for this control set is plotted in Fig. \ref{fig:DragFig3} as the red dash-dot-dot line. Its error rate is lower then both the standard Gaussian controls and the $Z$-only correction. This is the control procedure used in Ref. \cite{Chow2010a} and Ref. \cite{Lucero2010}, where it was referred to as simple DRAG and half derivative respectively. Here we will refer to this as simply $Y$-only correction.

The third control solution we consider is what we refer to as the optimal first order solution. This is achieved by minimizing the elements in $H_\mathrm{extra}^{(2)}(t)$ such that the second order corrections to the control fields are zero. From Eq. \eqref{eq:hextra2}, we require calculating the second order frame transformations. From Eqs. \eqref{eq:framecon} -- \eqref{eq:framecon2} these are 
\begin{eqnarray}
\label{eq:framecon2or2}
 s^{(2)}_{y,0,2}&=&-\frac{1}{2}t_g\Omega_G \lambda_1 (t_g\Omega_G+ s^{(1)}_{y,0,1}),\\
s^{(2)}_{x,0,2}&=&-\frac{1}{2}t_g\Omega_G \lambda_1 s^{(1)}_{x,0,1},\\
 s^{(2)}_{y,1,2}&=&-\lambda_1\dot{s}^{(1)}_{x,0,1},\\
s^{(2)}_{x,1,2}&=&\frac{1}{2}\lambda_1(2\dot\Omega_G+2\dot s^{(1)}_{y,0,1}-2t_g \Omega_G s^{(1)}_{z,1} + t_g\Omega_G s^{(1)}_{z,2}).\nonumber\\
\end{eqnarray}
 Using these expressions and requiring that the matrix elements of $H_\mathrm{extra}^{(2)}(t)$ are zero in the qubit subspace (and elements coupling to the qubit subspace are zero) results in ${s}^{(1)}_{x,0,1}(t)={s}^{(1)}_{z,1}(t)=0$ and ${s}^{(1)}_{y,0,1}(t)=-t_g\Omega_G(t)\lambda_1/4$.  Substituting these into Eq. \eqref{eq:controlfeilds} gives the controls fields
\begin{equation}
\begin{split}
\Omega_{x}(t) =& \Omega_G(t) ,\quad
\Omega_{y}(t)=  -\frac{\dot\Omega_G(t)\lambda_1}{2\Delta_2},\\
\delta (t)=& \frac{\Omega_G^2(t)}{4\Delta_2}[\lambda_1^2 -2\lambda_1].
\label{eq:env_from_s}
  \end{split}
\end{equation} This optimal first order solution is plotted in Fig. \ref{fig:DragFig3} as the solid purple line. Its error is substantially lower the the other first order correction methods.

Finally for completeness we also present the first order DRAG solution presented in Ref. \cite{Motzoi2009}. This occurs when we choose ${s}^{(1)}_{x,0,1}(t)={s}^{(1)}_{z,1}(t)=0$ and ${s}^{(1)}_{y,0,1}(t)=-\Omega_G(t)/2$, resulting in
\begin{equation}
\begin{split}
\Omega_{x}(t) = &  \Omega_G(t) ,\quad
\Omega_{y}(t) =  -\frac{\dot\Omega_G(t)}{\Delta_2},\quad\\
\delta(t) =& \frac{ \Omega_G^2(t)}{4\Delta_2}[\lambda_1^2-4].
  \end{split}
\end{equation} This solution can be intuitively derived from an interaction picture; however, there is nothing optimal about this choice. In Fig. \ref{fig:DragFig3},  the green dash-dot line shows how the error scales with this control set.  We see that for the first order solution, the error is larger than both the optimal and the $Y$-only correction methods. That is, the first order solution of Ref. \cite{Motzoi2009} is not optimal.

\subsubsection{Second order solution}

\begin{figure}[htbp]
\centering
\includegraphics[width=0.40\textwidth]{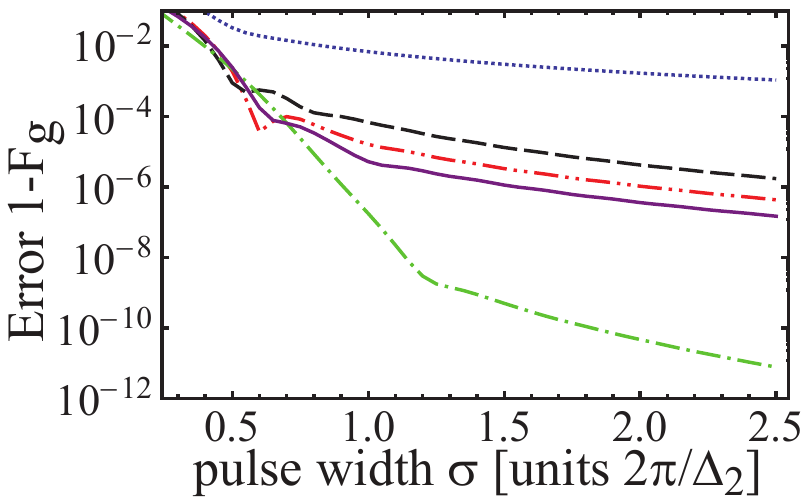}
\caption{(color online) Gate error for the implementation of a NOT gate in a $d=5$ SNO as a function of $\sigma$, with $t_g=4\sigma$, and a Gaussian shaped pulse. The blue dotted line is the zeroth order solution. The black dash line is the second order $Z$-only correction. The red dash-dot-dot line is the second order $Y$-only correction. The green dash-dot line is the second order correction from the controls presented in Ref. \cite{Motzoi2009}. The purple solid line is for the optimal first order correction.}\label{fig:DragFig4}
\end{figure}

The higher order solutions become impractical to solve in generality because the number of terms grows quickly with increasing order.  However, we can easily find the second and higher order corrections to the different first order solutions.  To do this, it is simplest to use a computer algebra system as the expressions for $H_\mathrm{extra}^{(2)}(t)$ and $H_\mathrm{extra}^{(3)}(t)$ are rather involved [see Eqs. \eqref{eq:hextra2} and \eqref{eq:hextra3}]. We find that
the corrections to the above four cases only change the $\Omega_{x}(t)$ field. In the $Z$-only case,  $\Omega_{x}(t)$ becomes
\begin{equation}
\Omega_{x}(t) =   \Omega_G(t) +\frac{\lambda_1^2\Omega_G^3}{8\Delta_2^2},
\end{equation} in the $Y$-only case, $\Omega_{x}(t)$ becomes
\begin{equation}
\Omega_{x}(t) =   \Omega_G(t) - \frac{\lambda_1^2(\lambda_1^2-4)\Omega_G^3}{32\Delta_2^2},
\end{equation} and the control set presented in Ref. \cite{Motzoi2009} gives
\begin{equation}
\Omega_{x}(t) =   \Omega_G(t) + \frac{(\lambda_1^2-4)\Omega_G^3}{8\Delta_2^2}.
\end{equation} To demonstrate these correction we plot in Fig. \ref{fig:DragFig4} the second order solutions. The line marking and colors are the same as in Fig. \ref{fig:DragFig3} with the exception that they now refer to second order solutions (all except the blue dotted line, which remains the zero order solution, and the purple solid line, which is the first order optimal solution). We see that the second order only makes small improvements to the $Y$-only (red dash-dot-dot) and $Z$-only (black dashed) first order solutions. Remarkably, the original DRAG scheme from Ref. \cite{Motzoi2009} (green dash-dot) is improved substantially when corrected to second order. It is for this reason that we argue this is the best solution for implementing a DRAG correcting pulse. 

We have not proven the DRAG solution to be optimal, and it seems likely that a better second order solution exists. To find the optimal solution, the matrix elements of $H_\mathrm{extra}^{(3)}(t)$ in the qubit subspace and the elements that couple to it must be minimized.  We have not computed this solution due to the complexity of $H_\mathrm{extra}^{(3)}(t)$ as well as the already outstandingly low error of the original DRAG solution.     

\subsection{Numerically optimized first order solutions}

\begin{figure}[htbp]
\centering
\includegraphics[width=0.40\textwidth]{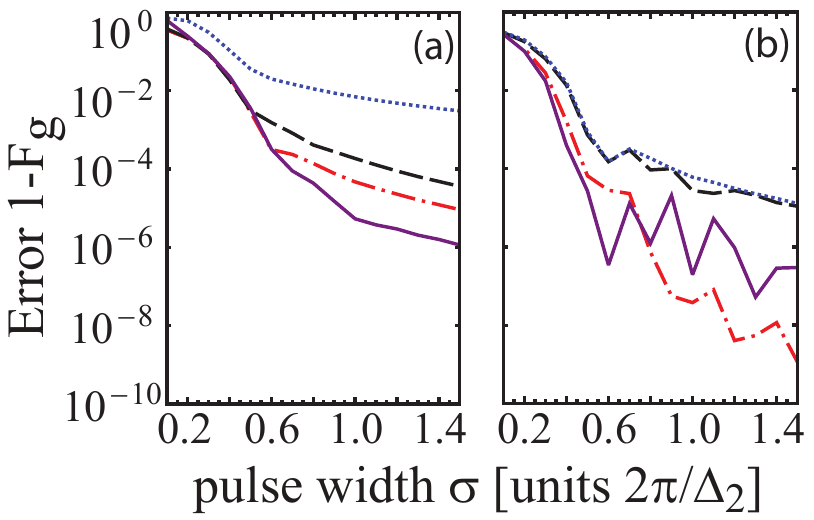}
\caption{(color online) Gate error for the implementation of a NOT gate in a $d=5$ SNO as a function of $\sigma$, with $t_g=4\sigma$, and a Gaussian shaped pulse. Panel (a) is for optimized first order solutions. The blue is for optimized $\alpha$ with $\beta=\gamma=\delta_0=0$ (zeroth order solution). The black dashed is for optimized $\alpha$ and $\gamma$ with $\beta=\delta_0=0$ ($Z$-only solution). Red dash-dot-dot is optimized $\alpha$ and $\beta$ with $\gamma=\delta_0=0$ ($Y$-only solution). Purple solid line is for optimized $\alpha$, $\beta$, and $\gamma$ with $\delta_0=0$ (optimal first order solution). Panel (b) is the same as (a) but with $\delta_0$ being optimized. }\label{fig:Numerical}
\end{figure}

Given the ease of implementing the first order solutions, in this section we consider the problem of numerically optimizing a value for the control fields with the following ansatz
\begin{equation}
\begin{split}
\Omega_{x}(t) =& \alpha\Omega_G(t) ,\quad
\Omega_{y}(t)=  -\beta\frac{\dot\Omega_G(t)}{\Delta_2},\\
\delta (t)=& \gamma\frac{\Omega_G^2(t)}{\Delta_2}+\delta_0, 
  \end{split}
\end{equation} where $\alpha$, $\beta$, $\gamma$  and $\delta_0$ are fit parameters. We consider a SNO with $\Delta_2=-2\pi$, $\lambda_{j-1}=\sqrt{j}$, $d=5$, and a control field given by Eq. \eqref{eq:gaussian} with $A=\pi$ and $t_g=4\sigma$ (same as before). In Fig. \ref{fig:Numerical} we plot the gate error as a function of $\sigma$ for different optimizations. The optimization procedure was done with Mathematica with a working precision of 10. In Fig. \ref{fig:Numerical} (a) we consider the case when $\delta_0=0$, and we find that optimizing the weighting of the control fields only improves the first order solutions slightly. This is expected as the second order solutions require different functional forms for the controls.
However, when we allow $\delta_0$ to be non-zero, we find some interesting results. For the numerical parameters considered, we find that implementing a time varying $\delta(t)$ ($\gamma\neq 0$) does not lead to any improvements. This is seen in Fig. \ref{fig:Numerical} (b) where we show that the error arising  from an optimized Gaussian with added constant detuning (blue dotted line) is approximately equal to the optimized $Z$-only correction with an added constant detuning (black dashed line).  Furthermore, the optimized $Y$-only correction with an added constant detuning (red dash-dot-dot) is approximately equal to the optimal first order solution with an added constant detuning (solid purple).  We also find that for the solutions with the derivative for the $Y$-control (solid purple and red dash-dot-dot) the gate error is much lower then in the other cases (blue dotted and black dashed). This gate error is approximately equal to those found with the second order corrections from Fig. \ref{fig:DragFig4} (green dash-dot line).  We conjecture from these numerics that the optimal DRAG-like solution can be obtained by applying a pulse to the $x$-axis (and its derivative to the $y$-axis) with a frequency that is not equal to the transition frequency of qubit.  

\section{Leakage extension}\label{sec:examples}

\begin{figure}[htbp]
\centering
\includegraphics[width=0.45\textwidth]{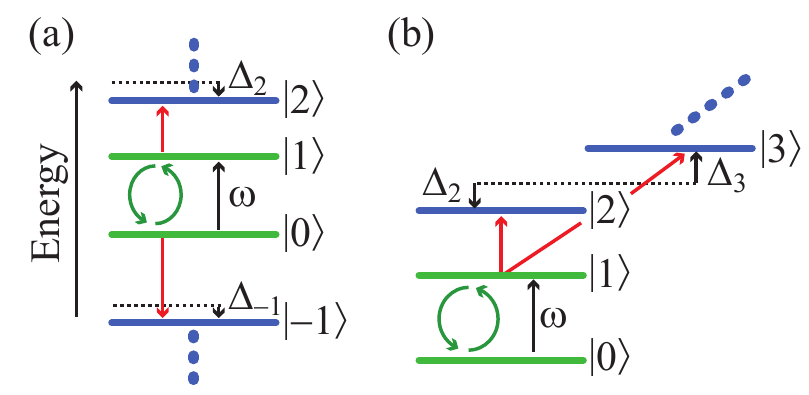}
\caption{(color online) Energy level diagrams for systems that can also be model by our theory. In (a) we consider the case when the qubit has leakage from both its $\ket{0}$ and $\ket{1}$ level. In (b) we consider the case when the qubit has leakage from its $\ket{1}$ level to more then one higher level.}\label{fig:Ext}
\end{figure}

In this section, we show that the DRAG technique can be applied to systems with more then one leakage transition. To show this we consider the two cases shown in Fig. \ref{fig:Ext}. We note that these are arbitrary examples, and the theory is more general then considered here [see appenix C]. 

\subsection{Leakage from both logical states}

\begin{figure}[htbp]
\centering
\includegraphics[width=0.40\textwidth]{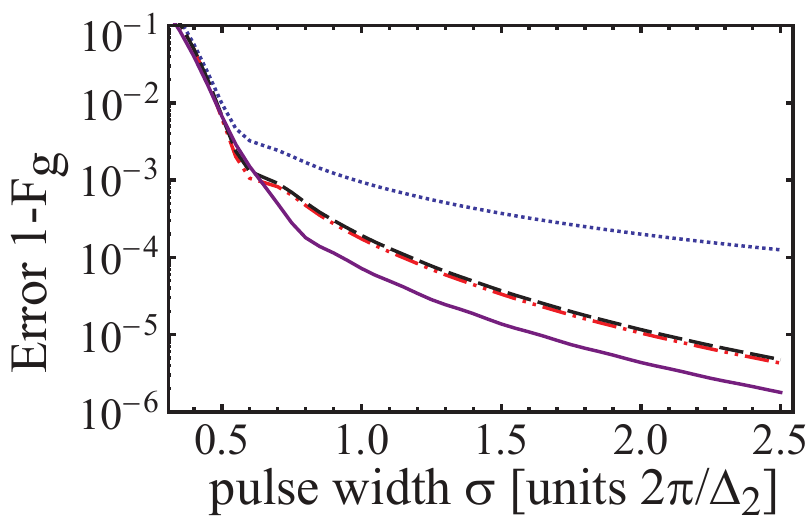}
\caption{(color online) Gate error for the implementation of a NOT gate when there is leakage above and below the qubit subspace. The system considered is explained in the text. The blue dotted line is the zeroth order solution. The black dash line is the first order $Z$-only correction. The red dash-dot-dot line is the first order $Y$-only correction. The purple solid line is for the optimal first order correction.}\label{fig:DragFig7}
\end{figure}

The first case we consider is a qubit defined in the intermediate states of an anharmonic oscillator [Fig.~\ref{fig:Ext} (a)].
This situation is important if the anharmonic oscillator is going to be used for qudit logic, as done in Ref. \cite{Neeley2009}, or for state tomography of the qudit, as done in Ref. \cite{Bianchetti2010}.
In this case, we rewrite the free and coupling Hamiltonian as
\begin{eqnarray}
    H_\mathrm{fr}&=&\sum_{j=-N}^{N}(j\omega+\Delta_j)\Pi_j,\\
H_\mathrm{ct}(t)&=&\mathcal{E}(t)\sum_{j=-N+1}^{N}\lambda_{j-1}\sigma^x_{j-1,j},
\label{eq:hamiltonianB}\end{eqnarray} where $N=(d-1)/2$ and again we take $\Delta_0=\Delta_1=0$. Moving to a interaction frame similar to Eq. \eqref{eq:rot} (the sum range is change to be consistent with  the above) we find a dimensionless rotating frame Hamiltonian equivalent to Eq. \eqref{eq:dimHam} with
\begin{eqnarray}
H_0 &=& \sum_{j=-N,\neq0,1}^{N} \frac{\Delta_j}{\Delta_2}\Pi_j,\\
H_z &=& \sum_{j=-N}^{N}j \Pi_j,\\
H_x &=& \sum_{j=-N+1}^{N}\lambda_{j-1} \sigma_{j-1,j}^x, \\
H_y &=& \sum_{j=-N+1}^{N} \lambda_{j-1} \sigma_{j-1,j}^y.
\end{eqnarray}
Using the results of appendix \ref{A:general}, the zero order dimensionless controls are the same as Eq. \eqref{eq:zerodim}, implying that the zeroth order controls are given by Eq. \eqref{eq:zero}.  To find the first order corrections we follow a similar procedure to Sec. \ref{sec:expansion}. The frame constraints from Eqs. \eqref{eq:framecon} and \eqref{eq:framecon2b}  become
\begin{eqnarray}
 s^{(n+1)}_{y,0,k}(t)&=&-\frac{\Delta_2}{2 \Delta_k}\mathrm{Tr}[{H}_\mathrm{extra}^{(n)}(t) \sigma^x_{0,k}]-\frac{\Delta_2\bar\Omega_x^{(n)}(t)\lambda_{-1}\delta_{k,-1}}{2\Delta_{-1}},\nonumber\\\\
s^{(n+1)}_{x,0,k}(t)&=&\frac{\Delta_2}{2 \Delta_k}\mathrm{Tr}[{H}_\mathrm{extra}^{(n)} (t)\sigma^y_{0,k}]+\frac{\Delta_2\bar\Omega_y^{(n)}(t)\lambda_{-1}\delta_{k,-1}}{2\Delta_{-1}},\nonumber\\
\end{eqnarray} while Eqs. \eqref{eq:framecon2a} and \eqref{eq:framecon2} remain the same.
With  $H_\mathrm{extra}^{(1)}(t)$ given by Eq. \eqref{eq:hextra1} we find the first order frame transformation to be
\begin{equation}
\begin{split}
  s^{(1)}_{x,0,k}(t) = &0,\quad s^{(1)}_{y,0,k}(t) =-\frac{\lambda_{-1}\Delta_{2} t_g\Omega_{G}(t)\delta_{k,-1}}{2\Delta_{-1}},\\ s^{(1)}_{x,1,k}(t) = &0,\quad s^{(1)}_{y,1,k} (t)= -\frac{\lambda_1 t_g\Omega_{G}(t)\delta_{k,2}}{2},
\end{split}
\end{equation}
 and the dimensionless first order control fields are
\begin{eqnarray}
\bar\Omega_{x}^{(1)} (t) &=&  2\dot{s}^{(1)}_{x,0,1} (t) ,\\
 \bar\Omega_{y}^{(1)}(t) &=&   2 \dot{s}^{(1)}_{y,0,1} (t)-s^{(1)}_{z,1} (t)t_g\Omega_G (t),\\
  \bar\delta^{(1)} (t) &= &\dot{s}^{(1)}_{z,1}(t) + 2 s^{(1)}_{y,0,1}(t)t_g\Omega_G(t) \\&&+ \frac{t_g^2\Omega_G^2(t)}{4}\textstyle{\left(\lambda_1^2 -\frac{\Delta_{2}}{\Delta_{-1}}\lambda_{-1}^2 \right).}
\end{eqnarray}
From this we find the control fields for $Z$-only correction are
\begin{equation}
\begin{split}
\Omega_{x}(t) =&  \Omega_G(t) ,\quad
\Omega_{y}(t) =  0,\quad\\
\delta(t) =& \frac{ \Omega_G^2(t)}{4}{\textstyle{ \left[\frac{\lambda_{1}^2}{\Delta_2}-\frac{\lambda_{-1}^2}{\Delta_{-1}}\right]}},
  \end{split}
\end{equation}
 the $Y$-only are
\begin{equation}
\begin{split}
\Omega_{x}(t) = &  \Omega_G(t) ,\quad
\Omega_{y}(t) =  -\frac{\dot\Omega_G(t)}{4}{\textstyle{\left[\frac{\lambda_{1}^2}{\Delta_2}-\frac{\lambda_{-1}^2}{\Delta_{-1}}\right],}}\quad\\
\delta(t) =& 0,
  \end{split}
\end{equation}
and the optimal first order control field corrections (after minimizing  $H_\mathrm{extra}^{(2)}(t)$) are
\begin{eqnarray}
\Omega_{x}(t) &= &  \Omega_G(t) ,\\
\Omega_{y}(t) &=&  -\frac{\dot\Omega_G(t)}{2\Delta_2}{\textstyle{\sqrt{\lambda_1^2+{{\frac{\Delta_2^2}{\Delta_{-1}^2}}}\lambda_{-1}^2}},}\\
\delta(t) &=&\frac{ \Omega_G^2(t)}{4\Delta_2}{\textstyle{\left[\lambda_1^2-\frac{\Delta_2^2}{\Delta_{-1}^2}\lambda_{-1}\right]}}.
\end{eqnarray}
To numerically demonstrate an improvement over the zeroth order solution we consider a SNO (of $d=6$) where we want to control the $2\rightarrow3$ transition. In this case, we relabel $j=2$ to 0 and so on, rescaling the coupling so that the new $0\rightarrow1$ transition is unity. This results in setting $\Delta_3=3\Delta_2$, $\Delta_{-1}=\Delta_2$, $\Delta_{-2}=3\Delta_2$, and $\lambda_0=1$, $\lambda_{1}=\sqrt{4/3}$, $\lambda_2=\sqrt{5/3}$ $\lambda_{-1}=\sqrt{2/3}$, $\lambda_{-2}=\sqrt{1/3}$. In Fig. \ref{fig:DragFig7}, the gate error for  implementing a NOT gate is shown as a function of $\sigma$ for the same Gaussian shaped pulse as considered in Sec. \ref{sec:error}. Here we observe that the DRAG technique improves the gate fidelities substantially when compared to the zeroth order solution and has error rates comparable to that of the case with only one leakage channel.

\subsection{Leakage from the excited state to more then one auxiliary level}

In the second case, shown in Fig. \ref{fig:Ext} (b), we consider a qubit made from the lowest 2 levels of a system which is coupled to many other transitions, all transitions having only a small energy cost (approximately $\Delta_2$).   This is an interesting example as it shows how this theory can be easily generalized.
In this case we rewrite the free and coupling Hamiltonians as
\begin{eqnarray}
    H_\mathrm{fr}&=&\omega\Pi_1+\sum_{j=2}^{d-1}(2\omega+\Delta_j)\Pi_j,\\
H_\mathrm{ct}(t)&=&\mathcal{E}(t)\bigg(\lambda_0\sigma^x_{0,1}+\sum_{j=2}^{d-1}\lambda_{j-1}\sigma^x_{1,j}\bigg).
\label{eq:hamiltonianC}\end{eqnarray} 
To eliminate the fast degrees of freedom we move to a rotating frame and make the standard rotating wave approximation. The procedure is similar to Sec. \ref{sec:Rot} with the replacement of  Eq. \eqref{eq:rot} by
\begin{equation}
 R(t)=\exp\left(-i \omega_d t \right)\Pi_1+\sum_{j=2}^{d-1}\exp\left(-i 2\omega_d t \right)\Pi_j.
\end{equation} This results in a dimensionless rotating frame Hamiltonian equivalent to Eq. \eqref{eq:dimHam} with
\begin{eqnarray}
H_0 &=& \sum_{j=2}^{d-1} \frac{\Delta_j}{\Delta_2}\Pi_j,\quad
H_z = \Pi_1+\sum_{j=2}^{d} 2\Pi_j,\nonumber\\
H_x &=& \lambda_0\sigma^x_{0,1}+\sum_{j=2}^{d-1}\lambda_{j-1}\sigma^x_{1,j}, \quad
H_y = \lambda_0\sigma^x_{0,1}+\sum_{j=2}^{d-1}\lambda_{j-1}\sigma^y_{1,j}.\nonumber
\end{eqnarray} Again, we find the zeroth order controls given by Eq. \eqref{eq:zero}.  To find the first order corrections, we follow a similar procedure as in Sec. \ref{sec:expansion}, finding that the frame constraints Eqs. \eqref{eq:framecon} and \eqref{eq:framecon2b} remain the same, but Eqs. \eqref{eq:framecon2a} and \eqref{eq:framecon2}  are changed to
\begin{eqnarray}
s^{(n+1)}_{y,1,k}(t)&=&-\frac{\Delta_2}{2 \Delta_k}\mathrm{Tr}[{H}_\mathrm{extra}^{(n)} (t)\sigma^x_{1,k}]-\frac{\bar\Omega_x^{(n)}(t)\lambda_{k-1}\Delta_2}{2\Delta_k},\nonumber\\\\
s^{(n+1)}_{x,1,k}(t)&=&\frac{\Delta_2}{2 \Delta_k}\mathrm{Tr}[{H}_\mathrm{extra}^{(n)}(t)\sigma^y_{1,k}]+\frac{\bar\Omega_y^{(n)}(t)\lambda_{k-1}\Delta_2}{2\Delta_k}. \nonumber\\
\end{eqnarray}
From the above with  $H_\mathrm{extra}^{(1)}(t)$ given by Eq. \eqref{eq:hextra1}, we find the first order frame transition to be
\begin{equation}
\begin{split}
  s^{(1)}_{x,0,k}(t) = &0,\quad s^{(1)}_{y,0,k}(t) =0,\\ s^{(1)}_{x,1,k}(t) = &0,\quad s^{(1)}_{y,1,k} (t)= -\frac{\lambda_{k-1} t_g\Omega_{G}(t)\Delta_{2}}{2\Delta_k}.
\end{split}
\end{equation}
 This gives the dimensionless first order control fields
\begin{equation}
\begin{split}
\bar\Omega_{x}^{(1)} (t) =&  2\dot{s}^{(1)}_{x,0,1} (t) ,\\
 \bar\Omega_{y}^{(1)}(t) =&   2 \dot{s}^{(1)}_{y,0,1} (t)-s^{(1)}_{z,1} (t)t_g\Omega_G (t),\\
  \bar\delta^{(1)} (t) = &\dot{s}^{(1)}_{z,1}(t) + 2 s^{(1)}_{y,0,1}(t)t_g\Omega_G(t) \\&+ \frac{t_g^2\Omega_G^2(t)\Delta_{2}}{4}{\textstyle{\left(\sum_{k=2}^{d-1}\frac{\lambda_{k-1}^2}{\Delta_{k}} \right).}}
  \end{split}
\end{equation}
From this, we find the control fields for the $Z$-only correction are 
\begin{equation}
\begin{split}
\Omega_{x}(t) = &  \Omega_G(t) ,\quad
\Omega_{y}(t) =  0,\quad\\
\delta(t) =& \frac{ \Omega_G^2(t)}{4}{\textstyle{\sum_{k=2}^{d-1}\frac{\lambda_{k-1}^2}{\Delta_{k}},}}
  \end{split}
\end{equation}
for the $Y$-only correction are
\begin{equation}
\begin{split}
\Omega_{x}(t) = &  \Omega_G(t) ,\quad
\Omega_{y}(t) =  -\frac{\dot\Omega_G(t)}{4}{\textstyle{\sum_{k=2}^{d-1}\frac{\lambda_{k-1}^2}{\Delta_{k}} ,}}\quad\\
\delta(t) =& 0,
  \end{split}
\end{equation}
and the optimal first order control field corrections (after minimizing  $H_\mathrm{extra}^{(2)}(t)$) are
\begin{equation}
\begin{split}
\Omega_{x}(t) = &  \Omega_G(t) ,\quad
\Omega_{y}(t) =  -\frac{\dot\Omega_G(t)}{2\Delta_2}{\textstyle{\sqrt{\sum_{k=2}^{d-1}\frac{\Delta_{2}^2\lambda_{k-1}^2}{\Delta_{k}^2}},}}\quad\\
\delta(t) =& \frac{ \Omega_G^2(t)}{4\Delta_2}{\textstyle{\left[\sum_{k=2}^{d-1}\frac{\Delta_{2}^2\lambda_{k-1}^2}{\Delta_{k}^2}-2\sqrt{\sum_{k=2}^{d-1}\frac{\Delta_{2}^2\lambda_{k-1}^2}{\Delta_{k}^2}} \right].}}
  \end{split}
\end{equation}
We note that these solutions are identical to the previous solutions with a single leakage channel where $\lambda_1=\tilde{\lambda}$,
\begin{equation}
\tilde{\lambda} \equiv {\textstyle{\sqrt{\sum_{k=2}^{d-1}\frac{\Delta_{2}^2\lambda_{k-1}^2}{\Delta_{k}^2}}}}.
\end{equation}

\begin{figure}[htbp]
\centering
\includegraphics[width=0.40\textwidth]{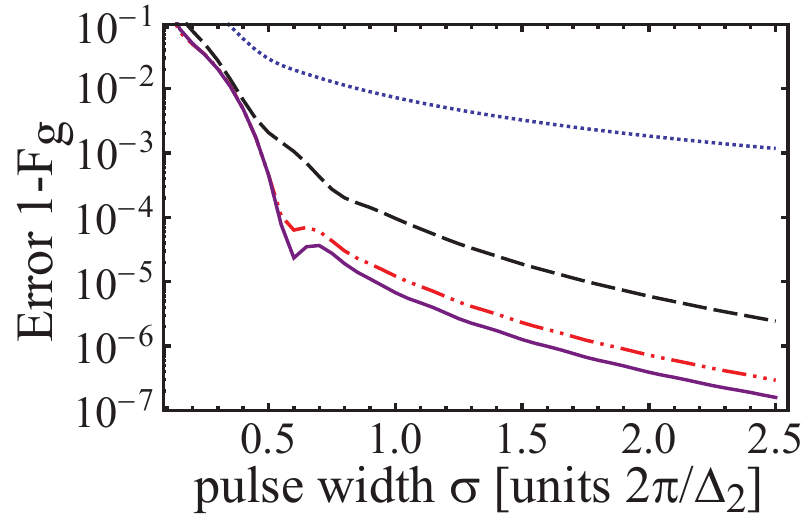}
\caption{(color online) Gate error for the implementation of a NOT gate when there is many leakage transitions for the excited state. The system considered is explained in the text. The blue dotted line is the zeroth order solution. The black dash line is the first order $Z$-only correction. The red dash-dot-dot line is the first order $Y$-only correction. The purple solid line is for the optimal first order correction.}\label{fig:DragFig8}
\end{figure}

To numerically demonstrate an improvement over the zeroth order solution, we consider the implementation of a NOT gate for a $d=6$ system with $\lambda_j=1$ for all $j$ and $\Delta_3=2\Delta_2$, $\Delta_{4}=3\Delta_2$, $\Delta_{5}=4\Delta_2$ (note these are different parameters from the anharmonic oscillator considered in Sec.~\ref{sec:expansion}). The results are plotted in Fig. \ref{fig:DragFig8}, where again, it is clearly seen that the DRAG technique improves the zeroth order solution.

\section{Conclusions}\label{sec:conclude}

In this paper we have presented a general technique for designing simple controls fields for single qubit unitary operations in weakly non-linear oscillators, which we refer to as the DRAG  (Derivative Removal by Adiabatic Gate) technique.  We first consider a qubit formed by the two lowest levels of an anharmonic oscillator with only the one photon transition elements being non-zero. In this system, the largest source of error for fast gates (small gate times) is leakage from the $\ket{1}$ state to the $\ket{2}$ state. Our technique provides a simple control methodology that perturbatively removes this error, thereby allowing high fidelity single qubit gates. The essential idea of this method is to apply a $y$-field that is proportional to the derivative of the original control pulse.  It contains the DRAG solution presented in Ref. \cite{Motzoi2009} as well as a large collection of control pulses that also correct for this leakage error, which, for example, require fewer control fields.  The lowest error obtained requires this derivative correction as well as a frequency shift.  Furthermore, we show that this methodology can be easily extended to other weakly non-linear oscillators with more then one leakage transition.  

In this paper, we considered Gaussian pulses due to their favorable spectral properties, however, our theory is independent of the initial form of the pulse shape.  A future research direction could be to find the optimal pulse shape for the leakage problem.  Furthermore, while the pulses found here are robust to first order in variations in the control parameters it would be interesting to search for pulses with higher order robustness properties.

\begin{acknowledgments}
We acknowledge, E. J. Pritchett for a carefull reading of the manuscript, B. Khani, J. M. Chow, L. DiCarlo, E. Lucero, J. M. Martinis and R. J. Schoelkopf for valuable discussions. J.M.G. was supported by CIFAR, DARPA/MTO QuEST program through a grant from AFOSR, Industry Canada, MITACS, MRI and NSERC. F.M., S.T.M., and
F.K.W.~were supported by NSERC through the discovery grants and
QuantumWorks. This research was also funded by the Office of the Director of National
Intelligence (ODNI), Intelligence Advanced Research Projects Activity
(IARPA), through the Army Research Office.  All statements of fact, opinion
or conclusions contained herein are those of the authors and should not be
construed as representing the official views or policies of IARPA, the ODNI,
or the U.S. Government.
\end{acknowledgments}

\appendix

\section{Typical functional forms for $\lambda_j$}\label{sec:lamb}

In this appendix, we discuss the different possible values $\lambda_j$ for the realistic models of direct driving \cite{Martinis2002} and the cavity \cite{Miller2005,Haroche2006} or circuit \cite{Blais2004,Wallraff2004} QED
architecture.
 In cases where the system is a SNO and is driven
directly, $\lambda_j$ is well approximated by the harmonic oscillator matrix elements, namely  $\lambda_j \approx \sqrt{j}$. This, for
example, occurs for the phase qubit when it is driven by a time-varying
bias current \cite{Martinis2002} and the transmon when it is driven by time-varying gate voltage \cite{Leek2009a}. Essentially, these systems are very nearly harmonic oscillators and the controls are almost proportional to the quadrature operator.

In the cavity or circuit QED
architecture, the anharmonic oscillator is coupled to a resonator and is control by driving the resonator far off-resonance. In this case, $\lambda_j$ can take on essentially any value. To see this, we start by writing the full Hamiltonian for the multi-level anharmonic oscillator and resonator as
\begin{equation}\label{eq:JC}
\begin{split}
H_\mathrm{JC} =& \omega_r a^\dagger a + \sum_{j=1}^{d-1}(j\omega+\Delta_j)\Pi_j \\&+ \sum_{j=1}^{d-1} g_{j-1,j}\left( \ket{j-1}\bra{j}a^\dagger+\ket{j}\bra{j-1}a\right),
\end{split}
\end{equation} again $\Delta_0=\Delta_1=0$. This is a generalized Jaynes-Cummings Hamiltonian \cite{Jaynes1963} with $\omega_r$ being the resonator frequency, and $g_{j,k}$ being the vacuum Rabi coupling for the $j$ to $k$ levels. Again we have assumed that the anharmonic oscillator only allows one photon transitions.

If $|\omega'_{j-1,j}-\omega_r|\gg |g_{j-1,j}|$ for all $j$ where $\omega'_{j-1,j}=\omega+\Delta_j-\Delta_{j-1}$ then diagonalization of Eq. \eqref{eq:JC} can be performed to lowest order in $g_{j-1,j}/(\tilde\omega_{j-1,j}-\omega_r)$ by the canonical transformation $H_\mathrm{JC}^\mathrm{D}= D^\dagger H_\mathrm{JC} D$ where
\begin{equation}\label{eq:D}D=\exp\left[\sum_{j=1}^{d-1} \frac{g_{j-1,j}}{\omega'_{j-1,j}-\omega_r}( a^\dagger \ket{j-1}\bra{j}-a \ket{j}\bra{j-1})\right]\end{equation}  to give
\begin{equation}
\begin{split}
H_\mathrm{JC}^D=&(\omega_r-\chi_{01}) a^\dagger a + \sum_{j=1}^{d-1}\left(j\omega+\Delta_j+\chi_{j-1,j}\right)\Pi_j\\&+\sum_{j=1}^{d-1}(\chi_{0,1}+\chi_{j-1,j}-\chi_{j,j+1})a^\dagger a\Pi_j.
\end{split}\label{eq:HJCD}
\end{equation} Here $\chi_{j-1,j}={g_{j-1,j}^2}/{(\omega'_{j-1,j}-\omega_r)}$ is the Lamb shift induced on the  anharmonic oscillator by the resonator and the last term in Eq. \eqref{eq:HJCD} is the ac-Stark shift \cite{Gambetta2006}.  Assuming that the dressed cavity is in vacuum  then Eq. \eqref{eq:HJCD} is well approximated by Eq. \eqref{eq:hamiltonian} with
\begin{eqnarray}\label{eqs:replace}
\omega&\rightarrow&\tilde\omega=\omega+\chi_{0,1}\\
\Delta_j&\rightarrow&\tilde\Delta_j=\Delta_j+\chi_{j-1,j}-j\chi_{0,1}
\end{eqnarray} and the tilde implies dressed values for the transition frequency and anharmonicity from the resonator.

As stated above, the control is usually through the resonator and is represented by the Hamiltonian
\begin{equation}H_\mathrm{dr}(t)=\varepsilon(t)(a+a^\dagger)\end{equation} which under the transformation Eq. \eqref{eq:D} becomes
\begin{equation}H^D_\mathrm{dr}(t)=\varepsilon(t)\left(a+a^\dagger + \sum_{j=1}^{d-1} \frac{g_{j-1,j}}{\omega'_{j-1,j}-\omega_r}\sigma_{j-1,j}^x\right). \label{eq:Hdr}\end{equation} Assuming that $\varepsilon(t)$ is a sinusoidal with a  frequency close to the qubit then Eq. \eqref{eq:Hdr} is well approximated by Eq. \eqref{eq:controlham} with
\begin{eqnarray}
\mathcal{E}(t)&=&\frac{g_{0,1}}{\omega'_{0,1}-\omega_r}\varepsilon(t)\\
\lambda_{j-1}&=&\frac{g_{j-1,j}(\omega'_{0,1}-\omega_r)}{g_{0,1}(\omega'_{j-1,j}-\omega_r)}.\label{eq:lambda}
\end{eqnarray}

\begin{figure}[htbp]
\centering
\includegraphics[width=0.40\textwidth]{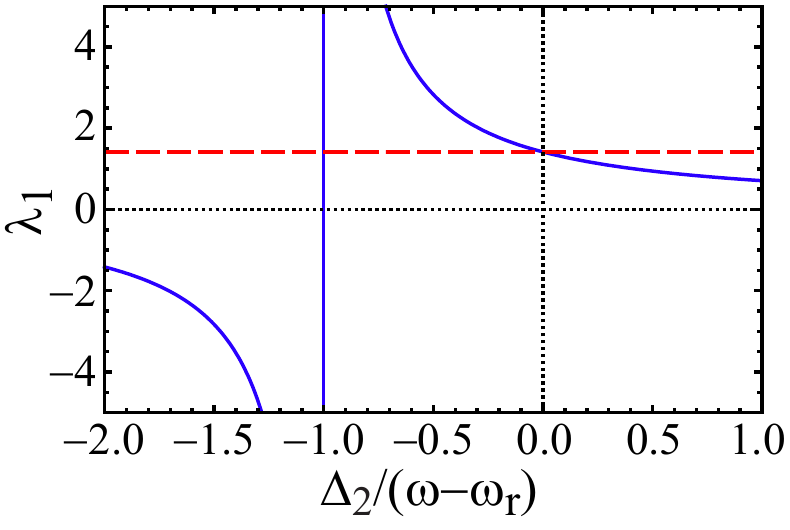}
\caption{(color online) Ratio of coupling strength of the $\ket{1}\rightarrow\ket{2}$ transition to the $\ket{0}\rightarrow\ket{1}$ transition as a function of anharmonicity for control through a resonator (solid  blue) and direct drive (dashed red). }\label{fig:Lambda}
\end{figure}

To demonstrate the functional form of $\lambda_1$, we will assume the anharmonic oscillator is in the SNO limit; $\omega'_{j-1,j}=\omega+(j-1)\Delta_2$, and $g_{j-1,j}=\sqrt{j}g_{0,1}$. In this case, Eq. \eqref{eq:lambda} becomes
\begin{equation}\label{eq:lambdaSNO}
\lambda_{j-1} = \frac{\sqrt{j}}{1+(j-1)\Delta_2/(\omega-\omega_r)},\end{equation} and is plotted in In Fig. \ref{fig:Lambda} as a function of the bare anharmonicity in units of $(\omega-\omega_r)$. Here, it is clearly seen that $\lambda_1$ is of not equal to $\sqrt{2}$ and actually changes sign at the point ($\omega-\omega_r=-\Delta_2$). This point invalidates the diagonalization for the second level and can not be treated under this model. Away from this point the values obtained from this model approximate the real situation. This was confirmed in Ref. \cite{Chow2010a} where the experimental value for $\lambda_1$ for the operation point used was found to agree with Eq. \eqref{eq:lambda}. However, we propose that $\lambda_j$ should be used as a fitting parameter in any experiment as effects such as higher modes of the resonator and higher order perturbation will result in additional corrections to this value \cite{Boissonneault2009}.

\section{Detuning control for fixed frequency qubits}\label{sec:det}

As a starting point for our control methods, we argue that the system can be explained by a Hamiltonian of the form displayed in Eq. \eqref{eq:rotHam}, with $\delta(t)$, $\Omega_x(t)$, and $\Omega_y(t)$ are independent control parameters. We argued that $\delta(t)$ is achieved by tuning the qubit frequency. Here, we show that it can also be achieved for fixed frequency qubits by changing the control field. In place of Eq. \eqref{eq:controls2}, we assume that input field is of the form
\begin{equation}\label{eq:controls2}
\begin{split}
\mathcal{E}(t)=&\Omega_x(t)\cos\left[\omega_d t + \int_0^t\delta(s)ds\right]\\&+\Omega_y(t)\sin\left[\omega_d t +\int_0^t\delta(s)ds\right]
\end{split}
\end{equation}

Now following the same procedure as in Sec. \ref{sec:Rot} with $\mathcal{E}(t)$ replaced by Eq. \eqref{eq:controls2} we can define a frame transformation $R(t)$
by
\begin{equation}
 R(t)=\sum_{j=1}^{d-1}\exp\left\{-i j\left[\omega_d t+ \int_0^t\delta(s)ds \right]\right\}\Pi_j,
\end{equation} and by using Eq. \eqref{eq:rotransf} we find an effective Hamiltonian of the form
\begin{equation}
 \begin{split}
H^R(t)=&\sum_{j=1}^{d-1}(j\delta(t)+j\delta_0+\Delta_j)\Pi_j\\& + \left[\frac{\Omega(t)}{2}e^{-i\omega_d t-i\int_0^t\delta(s)ds}+\mathrm{h.c.}\right]\\&\times\sum_{j=1}^{d-1}\lambda_{j-1}\left[\ket{j}\bra{j-1}e^{i\omega_d t+i\int_0^t\delta(s)ds}+\mathrm{h.c.}\right],\end{split}
\end{equation}  where $\Omega(t) = \Omega_x(t)+i\Omega_y(t)$. Now provided we can make the rotating wave approximation this simplifies to Eq. \eqref{eq:rotHam}.

Note Eq. \eqref{eq:controls2} may appear that is requires more control then is available but this is not true. Using simple trigonometric identities, this can be rewritten as
\begin{equation}\label{eq:controls3}
\begin{split}
\mathcal{E}(t)=&\Omega'_x(t)\cos\left[\omega_d t \right]+\Omega'_y(t)\sin\left[\omega_d t \right]
\end{split}
\end{equation}
where
\begin{eqnarray}
\Omega_x'(t) &=&\Omega_x(t)\cos\left[\int_0^t\delta(s)ds \right]+\Omega_y(t)\sin\left[\int_0^t\delta(s)ds \right]\nonumber\\\\
\Omega_y'(t) &=&\Omega_y(t)\cos\left[\int_0^t\delta(s)ds \right]-\Omega_x(t)\sin\left[\int_0^t\delta(s)ds \right].\nonumber\\
\end{eqnarray} This is known as phase ramping \cite{Patt1992}.

\section{General leakage removal}\label{A:general}

In this appendix, we give the general theory for finding adiabatic pulses for controlling subsystems of a larger system. The systems that will obey this theory are those in which transitions out of the subspace occur an energy cost $\Delta$. Furthermore, we assume this cost is the largest rate in the problem so that we  can form a perturbation expansion around $\epsilon=1/t_g\Delta$, where $t_g$ is the gate time. To be more specific, imagine a Hilbert space that can be partitioned into two subspaces, $\mathcal{H} = \mathcal{H}_{\rm{control}}\oplus \mathcal{H}_{\rm{leakage}}$, where we wish to perform some quantum gate solely in the control subspace, but with control Hamiltonians $H_m$ that act on the complete Hilbert space. That is, the general Hamiltonian for this system is
\be
H(t) = H_0 + \sum_m u_m(t) H_m,
\ee
 where we require that $H_0$ takes the form
 \be
H_0 = 0 \oplus \Delta \sum_k \frac{\Delta_k}{\Delta} \Pi_k.
\ee Here $\Pi_k=\ket{k}\bra{k}$ and we note that $H_0$ only has support on the leakage subspace.  Since, $\Delta$ is assumed to be large then $H_0$ is the dominant term in the Hamiltonian, and the leakage subspace is off-resonant.  It is less clear what constraints must be placed on the control Hamiltonians, $H_m$.  Certainly, if we wish to perform a general gate on the control subspace, then they must be controllable on that subspace, however, in order to completely eliminate leakage errors there are additional constraints, which we now determine.

The general protocol to determine these extra constraints arising from leakage is to first determine the Hamiltonian in the control subspace that implements the desired gate. That is, find the functions $h_l(t)$ of the Hamiltonian $H_\mathrm{sp}$
\be \label{eq:genHsp}
H_\mathrm{sp} = \sum_{l=1}^{d_c^2-1} h_l(t)B_l
\ee where $B_l$ is some orthogonal basis for operators in the control subspace of dimension $d_c$, that implement the desired gate. Next we define an effective Hamiltonian $H_\mathrm{eff}(t)$ in the complete space by Eq. \eqref{eq:Heff} with the only requirement on $A$ being that $A(0) = A(t_g) = \openone$ so that the frame transformation vanishes at the boundaries. The constraints on this effective Hamiltonian to be satisfied are
\begin{eqnarray}\label{eq:gconds}
&&\hspace{-5mm}\mathrm{Tr}[ H_\mathrm{eff}(t) B_l]=h_l(t),  \\
&&\hspace{-5mm}\bra{j} H_\mathrm{eff}(t)\ket{k}=0~ \mathrm{for}~0 \leq j < d_c~\mathrm{and}~d_c \leq k < d,\label{eq:gconds2}
\end{eqnarray} Here Eq. \eqref{eq:gconds} constrains the $H_\mathrm{eff}(t)$ to be equivalent to Eq. \eqref{eq:genHsp} and Eq. \eqref{eq:gconds2}  cancel the coupling between the control subspace and the leakage subspace. If theses constraints can be satisfied we have found a control set $\{u_m(t)\}$ that implements the desired gate with out leakage. In practice, this will not be possible but using the following procedure we can develop a perturbative approach.

The adiabatic transformation $A$ can be written as $A(t) = e^{-i S(t)}$ and we can write both $H(t)$ and the frame transformation $S(t)$ as power series in the small parameter $\epsilon$, as
\begin{equation}
 H(t) = \frac{1}{\epsilon}H_0+\sum_{n=0}^{\infty} \epsilon^n  H^{(n)}(t).
\end{equation} and
\be
S(t) = \sum_{n=1}^{\infty} S^{(n)}(t) \epsilon^l.
\ee Note the power expansion starts at $n=1$ as we restrict the transformation to be a perturbation in $\epsilon$.
Rewriting Eq.~\eqref{eq:Heff} in terms of $S$, as opposed to $A$, we obtain
\begin{equation}
\begin{split}
H_\mathrm{eff}(t) =& \sum_{j=0}^{\infty} \left( \textrm{ad}[i S(t)] \right)^j H(t) /j! \\
 &+ i \frac{\partial}{\partial t} \left(\sum_{j=0}^\infty  (i)^j S^j(t)/j! \right) \left(\sum_{k=0}^\infty (-i)^k S^k(t)/k! \right),
\end{split}
\end{equation}
where $\textrm{ad}[A]$ is the linear superoperator defined by $\textrm{ad}[A] B = [A,B] $ and we have used a Baker-Campbell-Hausdorff (BCH) and Taylor expansion for the first and last term respectively.  From this form of $H_\mathrm{eff}(t)$ we can isolate powers of $\epsilon$ and arrive at
\begin{equation}\label{eq:rec}
H_\mathrm{eff}^{(n)}(t)=H_\mathrm{extra}^{(n)}(t)+H^{(n)}(t) + i [S^{(n+1)}(t),H_0]
\end{equation} for $n\geq0$. Here $H_\mathrm{extra}^{(n)}(t)$ is a rather complicated expression which for the first few orders is
\begin{widetext}
\begin{eqnarray} \label{eq:hextra}
H_\mathrm{extra}^{(0)} (t)&=& 0 \\
H_\mathrm{extra}^{(1)}(t) &=&  i [S^{(1)}(t),H^{(0)}(t)]- [S^{(1)}(t),[S^{(1)}(t),H_0]]/2-\dot{S}^{(1)}(t).\label{eq:hextra1}\\
H_\mathrm{extra}^{(2)}(t) &=&  i [S^{(2)}(t),H^{(0)}(t)]+i [S^{(1)}(t),H^{(1)}(t)]- [S^{(1)}(t),[S^{(1)}(t),H^{(0)}(t)]]/2- [S^{(1)}(t),[S^{(2)}(t),H_0]]/2\nonumber\\&&
- [S^{(2)}(t),[S^{(1)}(t),H_0]]/2-i[{S}^{(1)}(t),[{S}^{(1)}(t),[{S}^{(1)}(t),H_0]]]/6+i[\dot{S}^{(1)}(t),{S}^{(1)}(t)]/2-\dot{S}^{(2)}(t).\label{eq:hextra2}
\\
H_\mathrm{extra}^{(3)}(t) &=&  i [S^{(3)}(t),H^{(0)}(t)]+i [S^{(2)}(t),H^{(1)}(t)]+i [S^{(1)}(t),H^{(2)}(t)]  - [S^{(1)}(t),[S^{(1)}(t),H^{(1)}(t)]]/2\nonumber
\\
&&- [S^{(1)}(t),[S^{(2)}(t),H^{(0)}(t)]]/2- [S^{(2)}(t),[S^{(1)}(t),H^{(0)}(t)]]/2-i[S^{(1)}(t),[S^{(1)}(t),[S^{(1)}(t),H^{(0)}(t)]]/6\nonumber
\\
&&- [S^{(1)}(t),[S^{(3)}(t),H_0]]/2- [S^{(3)}(t),[S^{(1)}(t),H_0]]/2-[S^{(2)}(t),[S^{(2)}(t),H_0]]/2\nonumber\\&&-i[{S}^{(1)}(t),[{S}^{(1)}(t),[{S}^{(2)}(t),H_0]]]/6
-i[{S}^{(2)}(t),[{S}^{(1)}(t),[{S}^{(2)}(t),H_0]]]/6-i[{S}^{(1)}(t),[{S}^{(2)}(t),[S^{(1)}(t),H_0]]]/6\nonumber\\&&+[{S}^{(1)}(t),[{S}^{(1)}(t),[{S}^{(1)}(t),[S^{(1)}(t),H_0]]]/24
+i[\dot{S}^{(1)}(t),{S}^{(2)}(t)]/2+i[\dot{S}^{(2)}(t),{S}^{(1)}(t)]/2\nonumber\\&&-[{S}^{(1)}(t),[\dot{S}^{(1)}(t),{S}^{(1)}(t)]/6-\dot{S}^{(3)}(t).\label{eq:hextra3}
\end{eqnarray}\end{widetext}
This expansion bears some formal similarity to the Schrieffer-Wolf/van-Vleck/adiabatic elimination expansion \cite{Shavitt80} which means that a compact closed expression to any order is unlikely to exist.
This expression for $H_\mathrm{eff}^{(n)}(t)$ is useful as it allows us to separate the free variables $H^{(n)}(t)$ and $S^{(n+1)}(t)$ for the order $n$ expression from those that are used to satisfy the conditions in Eqs. \eqref{eq:gconds} and \eqref{eq:gconds2} for order $n-1$, which only occur in the expression for $H_\mathrm{extra}^{(n)}(t)$. To see this more clearly we can used the above to rewrite Eq. \eqref{eq:gconds} for each order $n$ as
\be\label{eq:gcon}
\sum_m u_m^{(n)}\mathrm{Tr}[H_mB_l] = h_l^{(n)} -\mathrm{Tr}[H_\mathrm{extra}^{(n)}(t)B_l].
\ee for $0< l< d_c^2$ where  $h_l^{(n)}$ is the $n^\mathrm{th}$ order expression for $h_l$. Furthermore Eq. \eqref{eq:gconds} for each order $n$ becomes
\be\label{eq:gcon2}
S_{j,k}^{(n+1)}=i\frac{\Delta}{\Delta_k}  \bra{j} \left[H_\mathrm{extra}^{(n)}(t)+\sum_mu_m^{(n)} H_m \right]\ket{k}
\ee for $0 \leq j \leq d_c$ and $d_c \leq k < d$.
Here we have used that $H_0 $ has no weight in the control subspace. Since $H_\mathrm{extra}^{(n)}$ only depends on $S_{j,k}^{(n)}$, $H_0$, and $H^{(n-1)}$ then Eqs. \eqref{eq:gcon} and \eqref{eq:gcon2} are well formed.  The components of $S^{(n+1)}(t)$ that haven't yet been defined are free parameters in our theory.  Choosing a different functional form for the remaining matrix elements gives us the different forms of a DRAG solution.  It is necessary that the frame transformation vanish at $t= 0$ and $t=t_g$, which may restrict some of the freedoms we have in assigning value to the otherwise unconstrained elements of $S^{(n+1)}(t)$.  At each order of approximation, specifying the values of $H^{(n)}(t)$  and $S^{(n+1)}(t)$ fully determines $H_\mathrm{extra}^{(n+1)}(t)$ and so we can iterate this procedure indefinitely, though at some point the derivatives of $S$ that appear in $H_\mathrm{eff}(t)$ will likely become large, causing the adiabatic expansion to break down.

%\bibliography{Drag}
%\bibliographystyle{mybibtexstyle}

\end{document}